\documentclass[sigconf]{acmart}
\usepackage{multirow}
\usepackage{makecell}
\usepackage{subfigure}
\usepackage{pifont}

\AtBeginDocument{%
  }

\setcopyright{acmcopyright}
\copyrightyear{2023}
\acmYear{2023}
\setcopyright{acmlicensed}\acmConference[MM '23]{Proceedings of the 31st ACM International Conference on Multimedia}{October 29-November 3, 2023}{Ottawa, ON, Canada}
\acmBooktitle{Proceedings of the 31st ACM International Conference on Multimedia (MM '23), October 29-November 3, 2023, Ottawa, ON, Canada}
\acmPrice{15.00}
\acmDOI{10.1145/3581783.3612043}
\acmISBN{979-8-4007-0108-5/23/10}

\begin{document}

\title{Attributes Grouping and Mining Hashing for Fine-Grained Image Retrieval}

\author{Xin Lu}
\orcid{0009-0005-7669-1674}
\affiliation{
  \institution{School of Automation, Southeast University \& Key Laboratory of Measurement and Control of Complex Systems of Engineering, Ministry of Education, Southeast University}
  \city{Nanjing}
  \country{China}}
\email{lx_zjwf1218@outlook.com}

\author{Shikun Chen}
\orcid{0000-0002-0242-3133}
\affiliation{
  \institution{School of Automation, Southeast University \& Key Laboratory of Measurement and Control of Complex Systems of Engineering, Ministry of Education, Southeast University}
  \city{Nanjing}
  \country{China}}
\email{230198550@seu.edu.cn}

\author{Yichao Cao}
\orcid{0000-0003-2997-4012}
\affiliation{
  \institution{School of Automation, Southeast University \& Key Laboratory of Measurement and Control of Complex Systems of Engineering, Ministry of Education, Southeast University}
  \city{Nanjing}
  \country{China}}
\email{caoyichao@seu.edu.cn}

\author{Xin Zhou}
\orcid{0009-0006-5525-435X}
\affiliation{
  \institution{Nanjing Enbo Technology Co., Ltd.}
  \city{Nanjing}
  \country{China}}
\email{ustczhouxin@sina.com}

\author{Xiaobo Lu}
\authornote{Corresponding author.}
\orcid{0000-0002-7707-7538}
\affiliation{
  \institution{School of Automation, Southeast University \& Key Laboratory of Measurement and Control of Complex Systems of Engineering, Ministry of Education, Southeast University}
  \city{Nanjing}
  \country{China}}
\email{xblu2013@126.com}

\renewcommand{\shortauthors}{Xin Lu, Shikun Chen, Yichao Cao, Xin Zhou, \& Xiaobo Lu}

\begin{abstract}
In recent years, hashing methods have been popular in the large-scale media search for low storage and strong representation capabilities. To describe objects with similar overall appearance but subtle differences, more and more studies focus on hashing-based fine-grained image retrieval. Existing hashing networks usually generate both local and global features through attention guidance on the same deep activation tensor, which limits the diversity of feature representations. To handle this limitation, we substitute convolutional descriptors for attention-guided features and propose an Attributes Grouping and Mining Hashing (AGMH), which groups and embeds the category-specific visual attributes in multiple descriptors to generate a comprehensive feature representation for efficient fine-grained image retrieval. Specifically, an Attention Dispersion Loss (ADL) is designed to force the descriptors to attend to various local regions and capture diverse subtle details. Moreover, we propose a Stepwise Interactive External Attention (SIEA) to mine critical attributes in each descriptor and construct correlations between fine-grained attributes and objects. The attention mechanism is dedicated to learning discrete attributes, which will not cost additional computations in hash codes generation. Finally, the compact binary codes are learned by preserving pairwise similarities. Experimental results demonstrate that AGMH consistently yields the best performance against state-of-the-art methods on fine-grained benchmark datasets.
\end{abstract}


\maketitle

\section{Introduction}
Large-scale fine-grained image retrieval is an important research topic in multimedia \cite{li2018deep}. It has attracted increasing attention with the explosion of multimedia data. In contrast to fine-grained classification, the retrieval method faces an open-world problem with unlimited categories \cite{9609630}. The content-based fine-grained image retrieval aims to retrieve images from various sub-categories of a certain meta-category, which suffers from the subtle inter-class variations in similar sub-categories and intra-class variations caused by object postures \cite{9638382}. To meet the above challenges, the key is to explore the nuances embedded in the object attributes and utilize the high-level fine-grained features.

\begin{figure}[t]
\centering
\includegraphics[width=\linewidth]{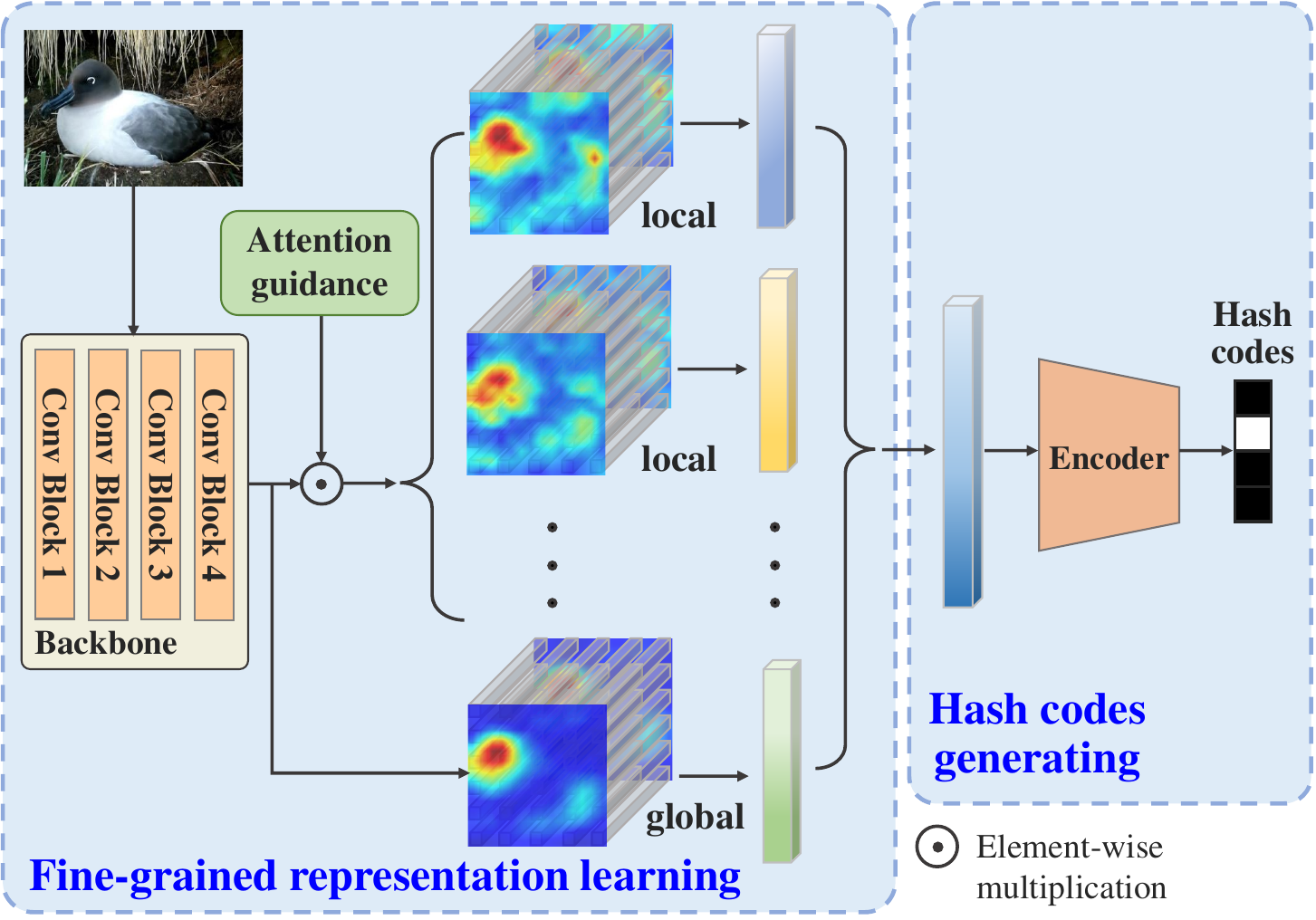}
\caption{The common framework of existing fine-grained hashing networks, which is mainly composed of fine-grained representation learning and hash codes generating. It relies on a set of attention guidance to generate diverse local and global features from the same intermediate features.}
\label{fig1}
\end{figure}

Recently, learning-to-hash methods \cite{7915742, 8432451} have achieved success in the tasks such as cross-modal hashing \cite{7867785, 9001235, 8331146} and are considered a promising solution for image retrieval. It maps the features from the deep convolutional layer to compact binary codes with high computing and storage efficiency. 
In practical applications, models usually need to retrieve the target category from a large number of similar objects belonging to the same meta-category. However, existing researches mainly explore the holistic representations of objects by hash codes, which are performed on coarse-grained datasets like NUS-WIDE \cite{chua2009nus} and CIFAR-10 \cite{krizhevsky2009learning}. It is more urgent to exploit fine-grained features and design a retrieval method that is more in line with real-world cases to obtain the expected data from massive images. Fine-grained hashing methods should be sensitive enough to discover the nuances of objects with similar appearances and support large-scale categories. SEMICON et al. \cite{cui2020exchnet, wei20212, shen2022semicon} propose to generate attentive local and global features and match fine-grained visual attributes, e.g., head color, tail color, living habits, with the bits of binary hash codes. The basic framework of these methods can be seen in Figure~\ref{fig1}. The intermediate features from backbone serve as the appetizer for both local and global features, which not only restricts the ability of feature representation but also leads to insufficient learning owing to the interaction of different features. 

\begin{figure*}[t]
\centering
\includegraphics[width=0.95\linewidth]{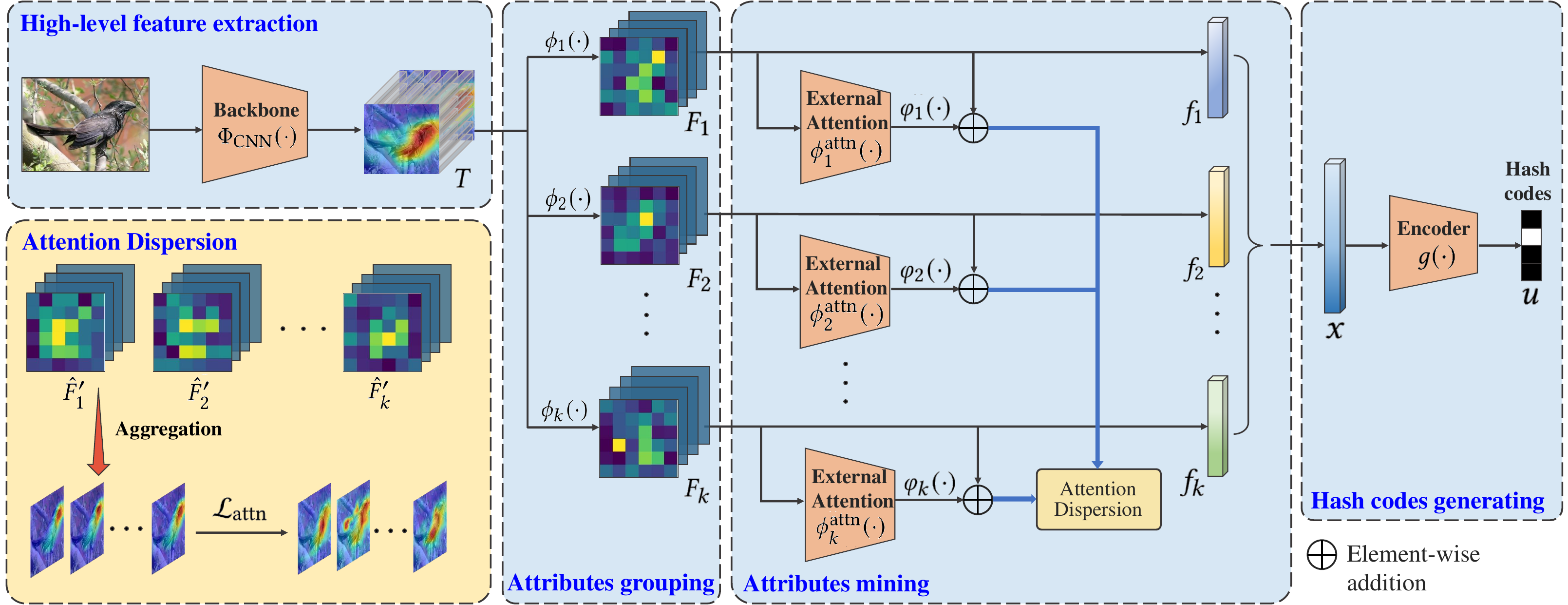}
\caption{The overall framework of AGMH. In the high-level feature extraction module, an image is fed into the backbone network to obtain base feature maps. Then, a set of convolution operations are conducted to group the object attributes and embed them in different descriptors. For the attributes mining module, the external attention mechanisms are attached to explore crucial attributes and learn dispersed attention. Finally, the compact binary codes are generated in the last module.}
\label{fig2}
\end{figure*}

In this paper, we argue that the various representations generated from the same intermediate features will inhibit the performance of hash learning. On the contrary, we compress the high-level feature extracted from the backbone network into multiple descriptors to learn various fine-grained attributes respectively. The slight adjustment can greatly improve the retrieval performance. Based on this discovery, a novel Attributes Grouping and Mining Hashing (AGMH) is proposed to learn the inter-class separable and intra-class compact hash codes. To avoid laborious and time-consuming annotations, we don’t leverage strong supervised information of part annotations and bounding boxes in the training phase but capture the discrete attributes automatically from learning. The framework of our proposed AGMH is shown in Figure~\ref{fig2}, which is composed of high-level feature extraction, attributes grouping, attributes mining, and hash codes learning. 

To be specific, the multiple descriptors can be utilized to learn object attributes in groups, which express the different parts of fine-grained objects in the representation space. To avoid learning the same attributes repeatedly, an Attention Dispersion Loss (ADL) is proposed to guide the descriptors to focus on diverse attributes. Considering that the direct application of ADL on descriptors may cause a potential negative impact on hash learning, we design a Stepwise Interactive External Attention (SIEA) to mine the implicit attributes embedded in each descriptor. It should be noted that the attention mechanism is integrated with each descriptor by skip connection, which is only exploited to improve the learning of discrete attributes but will not participate in hash learning. Therefore, the introduction of SIEA does not add extra computation in practical applications. It maintains the efficiency of our framework in large-scale multimedia data and promotes the accuracy of fine-grained retrieval. The main contributions of our work can be summarized as follows:
\begin{itemize}
\item We propose an attributes grouping and mining hashing to explore the diverse attributes in fine-grained objects through multiple convolutional descriptors. The framework can achieve superior performance in fine-grained image retrieval without bells and whistles.
\item We propose an attention dispersion loss to guide the descriptors to focus on diverse object attributes. Consequently, a more comprehensive feature representation of fine-grained objects can be obtained without handcrafted attention guidance.
\item A stepwise interactive external attention is designed to construct the correlations between different objects and attributes step by step. The attention mechanism helps to mine implicit attributes in descriptors and there is no testing complexity increment compared with the base network. 
\item Experimental results show that AGMH outperforms the state-of-the-art methods significantly on five fine-grained datasets, especially on short hash codes.
\end{itemize}

\section{Related Work}
\subsection{Content-based Fine-Grained Retrieval}
Fine-grained image retrieval can be divided into the content-based method \cite{wei2017selective, zheng2018centralized, zheng2019towards} and sketch-based method \cite{sain2020cross, pang2019generalising, ling2022multi} according to the type of query image. Among them, the content-based method is committed to retrieving images of various sub-categories with only subtle differences. It has underpinned a diverse set of real-world applications, especially online shopping, wildlife protection, and so on.

SCDA \cite{wei2017selective} is the earliest research of content-based fine-grain image retrieval with deep learning. It presents a pure unsupervised scheme by selecting and aggregating convolutional descriptors to localize main object and discard noisy background. CRL \cite{zheng2018centralized} greatly accelerate training compared to triplet loss and learn discriminative feature by a centralized global pooling. DGCRL \cite{zheng2019towards} employs a global ranking loss in the feature space and a Gram-Schmidt independent optimization in the center space to optimize the intra-class compactness and inter-class separability for fine-grained retrieval. Li et al. \cite{li2023fine} propose an attention context information constraints-based method for fine-grained image retrieval, which extracts useful local features from coarse to fine and models the internal semantic interactions of the learned local features. CNENet \cite{wang2022category} elaborately discovers category-specific nuances that contribute to category prediction, and semantically aligns these nuances grouped by sub-category without any additional prior knowledge, to directly emphasize the discrepancy among subcategories. 

In addition, aiming at the limitation of existing methods, several loss functions are proposed to improve the performance of fine-grained object retrieval. Hard Global Softmin Loss \cite{wang2022improved} is based on the global structure loss functions and hard mining strategy. SSCRL \cite{zeng2023sscrl} is a variant of softmax loss, named switched shifted softmax loss, to potentially reduce the overfitting phenomenon of the model. Piecewise Cross Entropy loss \cite{zeng2020fine} is proposed to enhance model generalization and promote the retrieval performance.

\subsection{Fine-Grained Deep Hashing}

Despite achieving good results, the above methods still face the dilemma that they are excessively time-consuming to search the nearest neighbor for given images in enormous data. Hence, hash learning \cite{andoni2008near, shen2017deep, jiang2018asymmetric}, which can map feature representations into compact binary codes, has been considered a promising scheme for fine-grained image retrieval. 

Ma et al. \cite{ma2020correlation} propose a correlation filtering hashing (CFH) method to learn discrete binary codes, which can adequately take advantage of the cross-modal correlation between the semantic information and the visual features for discriminative region localization. sRLH \cite{9638382} captures multifarious local discriminative information by locating the peaks of non-overlap sub-regions in the feature map and learns intra-class compact and inter-class separable hash codes. Recently, Chen et al. \cite{chen2022fine} raise three key issues that fine-grained hashing should address simultaneously and propose a novel Fine-graIned hashing (FISH) with a double-filtering mechanism and a proxy-based loss function. The above hashing methods take advantage of classification tasks to promote feature representation, which are distinct from the pure hash learning method and not emphasized in this paper.

DSaH \cite{jin2020deep} automatically mines discriminative regions by attention mechanism and generates semantic preserved hash codes. ExchNet \cite{cui2020exchnet} proposes attention constraints to obtain local features and aligns features by exchanging operations. A$^{2}$Net \cite{wei20212} establishes explicit correspondences between hash codes and visual attributes, which develops an encoder-decoder structure network to distill attribute-specific vectors from the appearance-specific representations. SEMICON \cite{shen2022semicon} localizes and restrains discriminative object parts to discover other complementary regions. The model is guided to exploit correlations across channels through interactive channel transformation in a stage-by-stage fashion. Motivated by the hashing networks with attention guidance, we adopt multiple descriptors to group and mine implicit attributes. The descriptors are generated from the high-level feature by convolution operations and guided by an elaborate loss function, which breaks the restriction of attention guidance in the above methods.

\section{Methodology}

\subsection{Overall Framework and Notations}
In fine-grained tasks, the representation of discriminative attributes is crucial to recognizing objects of sub-categories with subtle differences. To this end, AGMH learns attribute-aware features in two steps, i.e., related attributes grouping and crucial attributes mining. Given an input image $I$, the backbone network outputs deep activation tensor $T$ for preliminary representation, which embeds both object semantics and background information:
\begin{equation}
T=\Phi_{\mathrm{CNN}}(I)\in\mathbb{R}^{C\times{H}\times{W}}.
\end{equation}
Different from the previous work, we utilize a whole common network $\Phi_{\mathrm{CNN}}(\cdot)$ to extract features. Then, $k$ deep descriptors $\{F_{i}\}_{i=1}^k$ are generated by corresponding transforming networks $\{\phi_{i}{(\cdot)}\}_{i=1}^k$:
\begin{equation}
F_{i}=\phi_{i}(T)\in\mathbb{R}^{C'\times{H}\times{W}},
\end{equation}
where $\phi_{i}(\cdot)$ is a stack of convolution layers, which transforms semantic-specific features $T$ to attribute-level representation $F_{i}$ and reduces channel dimension to $C'$. 

Furthermore, an external attention mechanism $\phi_{i}^{\mathrm{attn}}(\cdot)$ is conducted on $F_{i}$ to mine attribute-aware local representations $\hat{F}_{i}$, and we add $\hat{F}_{i}$ with the original descriptor via skip connection:
\begin{gather}
\hat{F}_{i}=\phi_{i}^{\mathrm{attn}}(F_{i})\in\mathbb{R}^{C'\times{H}\times{W}}, \\
\hat{F}'_{i}=\sigma(F_{i}+\varphi_{i}(\hat{F}_{i})),\label{sc}
\end{gather}
where the $1\times1$ convolution layer $\varphi_{i}(\cdot)$ is employed to align $F_{i}$ and $\hat{F}_{i}$, and $\sigma$ indicates the rectified linear unit (ReLU) activation function. The discrete fine-grained attributes are learned under the guidance of elaborate loss. 

Eventually, we conduct global average pooling $\mathrm{GAP}(\cdot)$ to generate the feature $f_{i}\in\mathbb{R}^{C'}$ of part attributes and concatenate $\{f_{i}\}_{i=1}^k$:
\begin{gather}
f_{i}=\mathrm{GAP}(F_{i}), \\
x=[f_{1};f_{2};\cdots;f_{k}]_{\mathrm{cat}},
\end{gather}
where $[;]_{\mathrm{cat}}$ is concatenation operator, and $x\in\mathbb{R}^{kC'}$ indicates the holistic representation of the input image $I$. The $l$-length binary codes $u\in\{-1,+1\}^l$ are generated by hash network $g(\cdot)$:
\begin{equation}
u=g(x).
\end{equation}

\subsection{Stepwise Interactive External Attention}

In visual perception, people are accustomed to focusing on some specific regions and allocating information processing resources to important parts, which can be regarded as an attention mechanism. Similarly, attention mechanisms are of vital importance in deep feature representation for computer vision tasks. 

Self-attention \cite{vaswani2017attention} is the most representative work which captures the long-range dependency within a single sample, while it ignores the implicit relationship between different samples. External Attention \cite{guo2022beyond} memorizes the whole dataset with multi-layer perceptrons and introduces a multi-head mechanism to represent attention information in multiple subspaces. However, the tokens in multi-head external attention are interacted by shared memory units, which limit the representation of the correlations between different attributes. In this paper, we propose a stepwise interactive external attention (SIEA) to dig into the category-specific attributes implied in each descriptor and capture the relations between various objects and attributes. To save computation in testing, the attention mechanism does not participate in hash codes generation but improves the learning of discrete attributes in descriptors.

\begin{figure}[t]
\centering
\includegraphics[width=\linewidth]{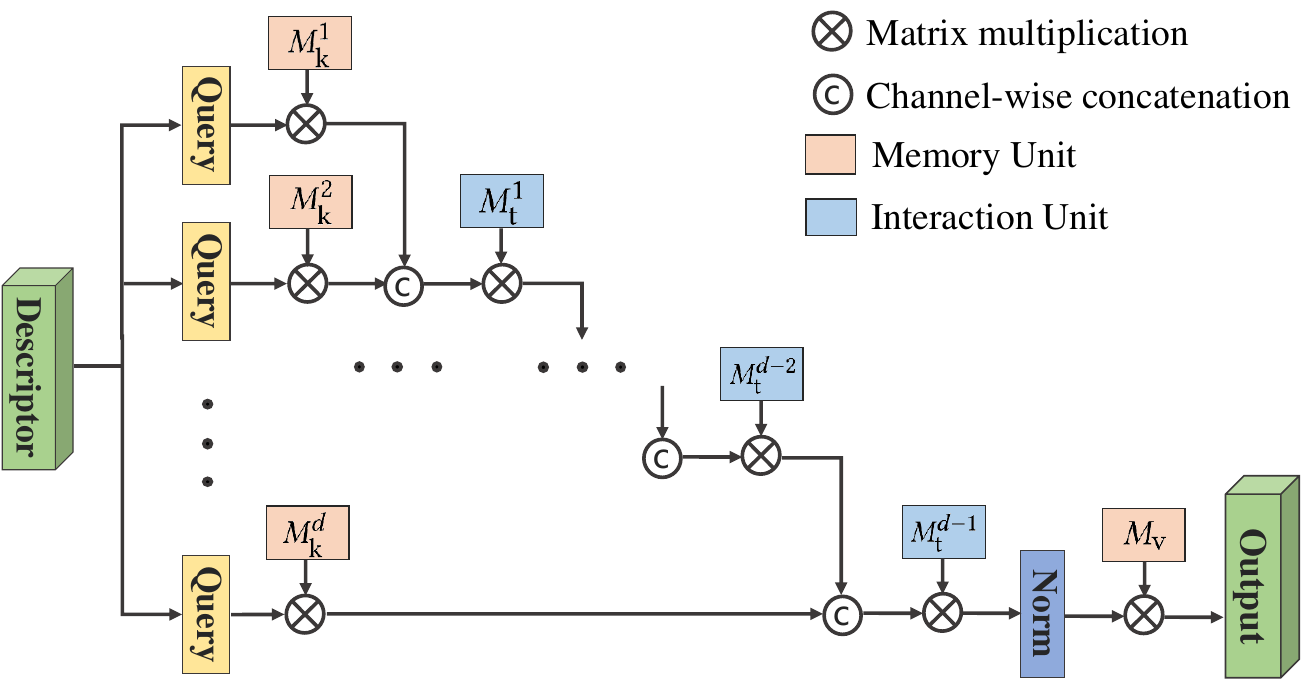}
\caption{The overview of Stepwise Interactive External Attention.}
\label{fig3}
\end{figure}

Figure~\ref{fig3} depicts our proposed SIEA. Given a deep descriptor $F_{i}\in\mathbb{R}^{C'\times{H}\times{W}}$, we first project it into a query matrix $Q_{i}\in\mathbb{R}^{N\times{C'}}$ by $1\times1$ convolution, transposition, and flattened in the spatial dimension, where $N=HW$ is the number of elements in a single channel. In external attention, the memory units of key and value, $M_{\mathrm{k}}\in\mathbb{R}^{S\times{C'}}$ and $M_{\mathrm{v}}\in\mathbb{R}^{S\times{C'}}$, are independent learnable parameters to generate attention maps and update the input features. Considering the multifarious attributes in each descriptor $F_{i}$, we adopt a set of memory units $\{M_{\mathrm{k}}^{j}\}_{j=1}^d$ to attend to different local discriminative information: 
\begin{equation}
G_{i}^{j}=Q_{i}{M_{\mathrm{k}}^{j}}^{\top},j=\{1,2,\cdots,d\}.
\end{equation}
It can be inferred that $G_{i}^{j}\in\mathbb{R}^{N\times{S}}$ expresses the similarity between the row vectors of $M_{\mathrm{k}}^{j}$ and $Q_{i}$. $\{G_{i}^{j}\}_{j=1}^d$ learn a variety of attention information in different representation spaces, which will boost the capability of external attention in the large-scale datasets. 

In addition, we introduce $d-1$ interaction units $M_{\mathrm{t}}\in\mathbb{R}^{S\times{2S}}$ to fuse $G_{i}^{j}$ step by step and compress redundant attention information. The stepwise interaction is formulated as:
\begin{equation}
P_{i}^{j}=
\begin{cases}
G_{i}^{1},  & j=1, \\
[P_{i}^{j-1};G_{i}^{j}]_{\mathrm{cat}}{M_{\mathrm{t}}^{j-1}}^{\top}, & j=\{2,\cdots,d\},
\end{cases}
\end{equation}
where $P_{i}^{j}$ represents the stepwise interacted attention information. The current attention information $G_{i}^{j}$ and the previous interaction result $P_{i}^{j-1}$ are concatenated in the channel dimension, thus $[P_{i}^{j-1};G_{i}^{j}]_{\mathrm{cat}}\in\mathbb{R}^{N\times{2S}}$. By utilizing multiple interaction units, we aggregate the attention information of various representation spaces into $P_{i}^{d}\in\mathbb{R}^{N\times{S}}$ and operate normalization on it:
\begin{gather}
\tilde{p}_{n,s}=\frac{\exp({p}_{n,s})}{\sum_{n=1}^{N}\exp({p}_{n,s})}, \\
\hat{p}_{n,s}=\frac{\tilde{p}_{n,s}}{\sum_{s=1}^{S}\tilde{p}_{n,s}},
\end{gather}
where ${p}_{n,s}$ is the point at position $(n,s)$ of $P_{i}^{d}$, and $\tilde{p}_{n,s}$ represents the attention value normalized by softmax in each channel. To reduce the sensitivity to feature scale, the point $\tilde{p}_{n,s}$ is performed with L1 norm in the column direction to obtain the final attention map $\hat{P}_{i}^{d}$.

Finally, we multiply $\hat{P}_{i}^{d}$ with the memory unit of value $M_{\mathrm{v}}$ to obtain the attributes-aware representation for the $i$-th descriptor:
\begin{equation}
\hat{F}_{i}=\hat{P}_{i}^{d}M_{\mathrm{v}}.
\end{equation}
To integrate $\hat{F}_{i}$ with $F_{i}$ as Equation (\ref{sc}), $\hat{F}_{i}$ will be restored to a three-dimensional map consistent with $F_{i}$, thus the attentive descriptor $\hat{F}_{i}\in\mathbb{R}^{C'\times{H}\times{W}}$.

\begin{figure}[t]
\centering
\includegraphics[width=\linewidth]{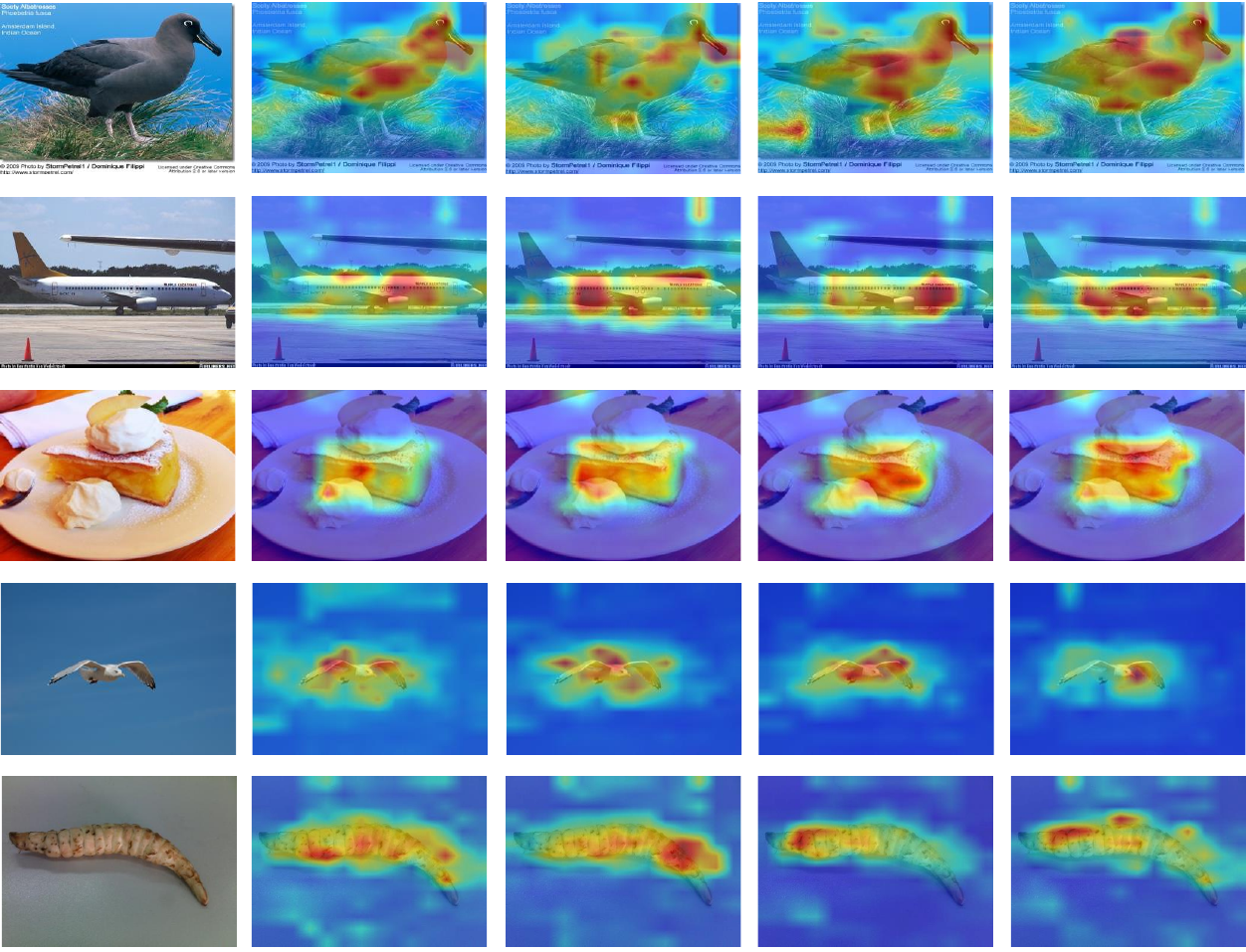}
\caption{Visualization of aggregation maps of attentive descriptor groups, which are generated from the objects in various fine-grained datasets. The focused regions are located in different local attributes.}
\label{fig4}
\end{figure}

\subsection{Attention Dispersion Loss}
Through SIEA, we attend to the critical attributes implied in each descriptor. The descriptors are learned independently before being mapped to Hamming space. No explicit relationship of spatial attention has been constructed between different descriptors in the learning phase. Some salient attributes will be repeatedly concerned, while some subtle attributes are prone to be overlooked, limiting the representation capability of multiple descriptors.

To this end, we design a novel attention dispersion loss (ADL) for the representations of multifarious attributes, which is adapted to our multi-branch structure. SCDA \cite{wei2017selective} adds up the activation tensor through depth direction to generate an aggregation map, which can effectively localize objects. For fine-grained descriptors, treating each channel equally will reduce the response variance of attributes in the aggregation map. To obtain discrete attribute responses, we select the maximum response corresponding to each spatial position as the aggregation value:
\begin{equation}
A_{i}({h,w})=\max(\hat{F}'_{i}({h,w})),
\end{equation}
where $A_{i}\in\mathbb{R}^{{H}\times{W}}$ represents the aggregation map of $\hat{F}'_{i}$. $a=A_{i}(h,w)$ is the maximum value of activation vector $\hat{f}\in\mathbb{R}^{C'\times{1}}$ at the position $(h,w)$ of $\hat{F}'_{i}$. It can be inferred that $A_{i}$ enhances the discrepancy between the focused attributes and other regions, which is more sensitive to critical attributes. 

To avoid excessive attention to some specific attributes, we exploit attention dispersion loss $\mathcal{L}_{attn}$ to motivate the attentive descriptors to focus on different attributes:
\begin{gather}
\tilde{A}_{i}(h,w)=\frac{\exp(A_{i}(h,w))}{\sum_{h=1}^{H}\sum_{w=1}^{W}\exp(A_{i}(h,w))}, \\
\mathcal{L}_{\mathrm{attn}}(I)=\frac{\sum_{i=1}^{k}\sum_{j=i+1}^{k}\left<\tilde{A}_{i},\tilde{A}_{j}\right>}{k(k+1)/2},\label{La}
\end{gather}
where $k$ is the number of descriptors, and $\left<\cdot,\cdot\right>$ denotes inner product. The activation responses on $A_{i}$ are normalized into probability distribution by softmax, and $\mathcal{L}_{attn}$ is formulated as the average inner product of all normalized map pairs $\tilde{A}_{i}$ and $\tilde{A}_{j}$. As can be seen from Figure~\ref{fig4}, ADL enlarges the discrepancy between the distributions of focused attributes in different descriptors.

\subsection{Hash Codes Learning}
Based on the generated descriptors $\{F_{i}\}_{i=1}^k$, we obtain the holistic representation $x\in\mathbb{R}^{kC'}$ as aforementioned and feed them into hash learning module afterward. To be specific, a linear encoder paradigm $W\in\mathbb{R}^{l\times{kC'}}$ is built to project $x$ as an internal latent representation $v\in\mathbb{R}^{l}$. The hash codes $u\in\{-1,+1\}^l$ of the input image $I$ are activated by the sign function $\mathrm{sign}(\cdot)$:
\begin{gather}
v=Wx, \\
u=\mathrm{sign}(v).\label{sv}
\end{gather}

Concerning $m$ internal latent representations $\{p_{t}\}_{t=1}^{m}$ of a database, the corresponding hash codes can be expressed as $\{z_{t}\in\{-1,+1\}^{l}\}_{t=1}^{m}$:
\begin{equation}
z_{t}=\mathrm{sign}(p_{t}).
\end{equation} 
To preserve the pairwise similarity of query codes and database codes, we perform hash codes learning through the following objective function:
\begin{equation}
\mathcal{L}_{\mathrm{hash}}(I)=\sum_{t=1}^{m}(\left<u,z_{t}\right>-lS_{t})^2,
\end{equation}
where $S\in\{-1,+1\}^{m}$ represents the pairwise supervised information, and $l$ is the length of binary hash codes. However, the gradient can not be back-propagated owing to the sign function. Hence, we substitute $\mathrm{sign}(\cdot)$ for $\mathrm{tanh}(\cdot)$ in Equation (\ref{sv}) to relax the whole optimization process, and the loss of hash learning can be reformulated as:
\begin{gather}
\tilde{u}=\mathrm{tanh}(v), \\
\mathcal{L}_{\mathrm{hash}}(I)=\sum_{t=1}^{m}(\left<\tilde{u},z_{t}\right>-lS_{t})^2+\alpha(\tilde{u}-u)^2,\label{Lh}
\end{gather}
where the last item of $\mathcal{L}_{hash}$ is the quantization loss to make $\tilde{u}$ close to $\{-1,+1\}^l$. 

\begin{table*}[t]
\caption{The retrieval accuracy (\% mAP) of AGMH and compared fine-grained hashing methods (based on ResNet-50) on four benchmark datasets. The best results under different hash bits are highlighted in bold.}
\label{tab1}
\begin{tabular*}{0.98\linewidth}{c|c|ccccccccc|c}
\toprule
\makebox[0.12\textwidth][c]{Datasets} & \makebox[0.05\textwidth][c]{\# bits} & \makebox[0.05\textwidth][c]{ITQ} & \makebox[0.05\textwidth][c]{SDH} & DPSH & HashNet & ADSH & ExchNet & A$^{2}$-Net & FCAENet & SEMICON & \makebox[0.065\textwidth][c]{Ours} \\
\hline  
\multirow{4}*{CUB200-2011} & 12 & 6.80 & 10.52 & 8.68 & 12.03 & 20.03 & 25.14 & 33.83 & 34.76 & 37.76 & \textbf{56.42} \\
 & 24 & 9.42 & 16.95 & 12.51 & 17.77 & 50.33 & 58.98 & 61.01 & 67.67 & 65.41 & \textbf{77.44} \\
 & 32 & 11.19 & 20.43 & 12.74 & 19.93 & 61.68 & 67.74 & 71.61 & 73.85 & 72.61 & \textbf{81.95} \\
 & 48 & 12.45 & 22.23 & 15.58 & 22.13 & 65.43 & 71.05 & 77.33 & 80.14 & 79.67 & \textbf{83.69} \\
\hline
\multirow{4}*{Aircraft} & 12 & 4.38 & 4.89 & 8.74 & 14.91 & 15.54 & 33.27 & 42.72 & 43.92 & 49.87 & \textbf{71.64} \\
 & 24 & 5.28 & 6.36 & 10.87 & 17.75 & 23.09 & 45.83 & 63.66 & 75.46 & 75.08 & \textbf{83.45} \\
 & 32 & 5.82 & 6.90 & 13.54 & 19.42 & 30.37 & 51.83 & 72.51 & 81.61 & 80.45 & \textbf{83.60} \\
 & 48 & 6.05 & 7.65 & 13.94 & 20.32 & 50.65 & 59.05 & 81.37 & 81.34 & 84.23 & \textbf{84.91} \\
\hline
\multirow{4}*{Food101} & 12 & 6.46 & 10.21 & 11.82 & 24.42 & 35.64 & 45.63 & 46.44 & 44.97 & 50.00 & \textbf{62.59} \\
 & 24 & 8.20 & 11.44 & 13.05 & 34.48 & 40.93 & 55.48 & 66.87 & 76.56 & 76.57 & \textbf{80.94} \\
 & 32 & 9.70 & 13.36 & 16.41 & 35.90 & 42.89 & 56.39 & 74.27 & 81.37 & 80.19 & \textbf{82.31} \\
 & 48 & 10.07 & 15.55 & 20.06 & 39.65 & 48.81 & 64.19 & 82.13 & 83.14 & 82.44 & \textbf{83.21} \\
\hline
\multirow{4}*{VegFru} & 12 & 3.05 & 5.92 & 6.33 & 3.70 & 8.24 & 23.55 & 25.52 & 21.76 & 30.32 & \textbf{43.99} \\
 & 24 & 5.51 & 11.55 & 9.05 & 6.24 & 24.90 & 35.93 & 44.73 & 50.36 & 58.45 & \textbf{68.05} \\
 & 32 & 7.48 & 14.55 & 10.28 & 7.83 & 36.53 & 48.27 & 52.75 & 67.46 & 69.92 & \textbf{76.73} \\
 & 48 & 8.74 & 16.45 & 9.11 & 10.29 & 55.15 & 69.30 & 69.77 & 79.76 & 79.77 & \textbf{84.49} \\
\bottomrule   
\end{tabular*}
\end{table*}

By combining Equation (\ref{La}) and (\ref{Lh}), the proposed framework AGMH can be optimized as:
\begin{equation}
\begin{aligned}
\min_{\Theta}\mathcal{L}(\mathcal{I}) & =\sum_{r\in\Omega}\left[\mathcal{L}_{\mathrm{hash}}(\mathcal{I}_{r})+\beta\mathcal{L}_{\mathrm{attn}}(\mathcal{I}_{r})\right] \\
& =\sum_{r\in\Omega}\sum_{t\in\Gamma}(\left<\tilde{u}_{r},z_{t}\right>-lS_{r,t})^2+\alpha(\tilde{u}_{r}-u_{r})^2 \\
& +\beta\sum_{r\in\Omega}\frac{\sum_{i=1}^{k}\sum_{j=i+1}^{k}\left<\tilde{A}_{i}^{r},\tilde{A}_{j}^{r}\right>}{k(k+1)/2} \\
& \mathrm{s.t.} \quad u_{r}\in\{-1,+1\}^{l},z_{t}\in\{-1,+1\}^{l},\label{Li}
\end{aligned}
\end{equation}
where $\Theta$ denotes all the parameters in AGMH, including $\Phi_{\mathrm{CNN}}$, $W$, and $\{\phi_{i}(\cdot),\phi_{i}^{\mathrm{attn}}(\cdot),\varphi_{i}(\cdot)\}_{i=1}^k$. $\alpha$ and $\beta$ are hyper-parameters as the trade-off. $\Gamma$ is denoted as the indices of database points. $\mathcal{I}=\{I_{r}\}_{r=1}^{n}$ is a set of query images randomly sampled from the database, and $\Omega\subseteq{\Gamma}$ indicates the indices of the query set. The pairwise supervised information $S_{r,t}\in\{-1,+1\}^{n\times{m}}$.

\subsection{Out-of-Sample Extenstion}
After the training phase, AGMH can be employed to generate hash codes for a query image $\mathcal{I}_{\mathrm{q}}$ that has not been learned. We can extract a holistic representation $x_{\mathrm{q}}$ from $\mathcal{I}_{\mathrm{q}}$ and generate the corresponding binary codes $u_{\mathrm{q}}$:
\begin{equation}
u_{\mathrm{q}}=\mathrm{sign}(Wx_{\mathrm{q}}).
\end{equation}

\section{EXPERIMENTS}
\subsection{Datasets}
In this section, we carry out a series of ablation studies and comparisons with the state-of-the-art methods on five fine-grained datasets, i.e., CUB200-2011 \cite{wah2011caltech}, Aircraft \cite{maji2013fine}, Food101 \cite{bossard2014food}, VegFru \cite{hou2017vegfru}, and Stanford Dogs \cite{khosla2011novel}. The official training and testing sets are utilized as retrieval and query sets in the evaluation.

CUB200-2011 is one of the most popular datasets for fine-grained image recognition. It contains 11,788 images of 200 bird categories, which is officially split into 5,994 training images and 5,794 test images. Aircraft is a set of 100 aircraft variants with 6,667 training images and 3,333 test images. In addition, the remaining fine-grained datasets are of large scale. Food101 consists of 101,000 images equally distributed among 101 kinds of foods. In each category, 250 test images have been checked manually, while there are certain noisy data in 750 training images. VegFru includes not only 200 kinds of vegetables but also 92 kinds of fruits. The images in VegFru are divided into 29200 training samples, 14600 validation samples and 116931 test samples. Standford Dogs is another dataset for dogs. It contains 120 dog species with 20580 images, where 100 images per class are selected to form the training set and the remaining images are for test.

\subsection{Implementation Details}
To compare with the SOTA fine-grained hashing methods, we adopt the same training settings as ExchNet, A$^{2}$-Net, and SEMICON. To be specific, all experiments are conducted on two NVIDIA Geforce RTX 3090 GPUs and two 2080 GPUs. In each training epoch, we sample 2000 images randomly for CUB200-2011, Aircraft, Food101, and Stanford Dogs, while 4000 stochastic images are sampled for VegFru. Following SEMICON, AGMH is trained 30 epochs per iteration with Stochastic Gradient Descent (SGD). The batch size is set as 64 with an input resolution of $224\times224$. The iteration time for CUB200-2011, Aircraft, and Stanford Dogs is 40, which increases to 50 for the other large-scale datasets. The initial learning rate is 0.001 and will be divided by 10 at 40-th iteration. The hyper-parameters $\alpha$ and $\beta$ in Equation (\ref{Li}) are set as 1 and 0.5 to balance the contribution of different losses. Besides, we set the number of descriptors $k$ as 6 and the channel dimension of descriptor $C'$ as 512 to group various fine-grained attributes. The number and input dimension of memory units, i.e., $d$ and $S$, are set as 4 and 128 to make a trade-off between efficiency and accuracy.

ExchNet et al. employ the first three stages of ResNet-50 as their backbone and the fourth stage without downsampling to generate attentive global and local features. Similarly, we extract high-level features from a complete ResNet-50. While in comparison with classification-promoted networks, AGMH is based on ResNet-18 just like sRLH, FISH, etc. 

\subsection{Comparison with the SOTA methods}
\textbf{Main results.} To prove the superiority of our AGMH, we compare it with existing hashing-based retrieval models. Among them, ITQ \cite{gong2012iterative}, SDH \cite{7298598}, DPSH \cite{li2015feature}, HashNet \cite{cao2017hashnet}, and ADSH \cite{jiang2018asymmetric} are coarse-grained methods, where ITQ and SDH are not based on deep learning. ExchNet \cite{cui2020exchnet}, A$^{2}$-Net \cite{wei20212}, FCAENet \cite{zhao2022feature}, and SEMICON \cite{shen2022semicon} are the most advanced hashing methods aiming at large-scale fine-grained retrieval.

Table~\ref{tab1} summarizes the mean average precision (mAP) of different retrieval methods on CUB200-2011, Aircraft, Food101, and VegFru, where the experiments are conducted on four hash code lengths, e.g., 12, 24, 32, 48. According to the main results, we can observe that our AGMH surpasses the existing hashing networks by large margins, especially on short hash bits. To be specific, compare with the state-of-the-art method SEMICON on 12-bit hash codes, the mAP of AGMH improves by 18.66\%, 21.77\%, 12.59\%, and 13.67\% on the benchmark datasets respectively. With the increase of hash bits, the performance gaps between hashing methods gradually decrease as the long codes can guide the learning of discriminative features. While our AGMH, without any sophisticated modules, still outperforms other fine-grained hashing networks. It achieves mAP results with 9.34\% and 4.02\% improvements over SEMICON of 32-bit and 48-bit experiments on CUB200-2011 as well as 6.81\% and 4.72\% improvements on the large-scale dataset VegFru.

\begin{figure}[t]
\centering
\includegraphics[width=0.80\linewidth]{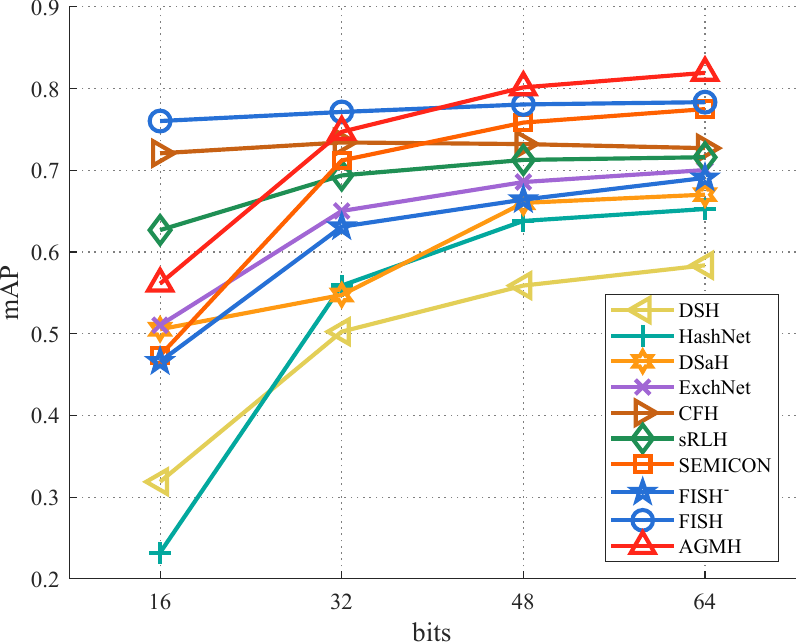}
\caption{The mAP results on CUB200-2011 with code length varying from 16 to 64. FISH$^{-}$ indicates the FISH trained without classification losses.}
\label{fig5}
\end{figure}

\noindent\textbf{Comparison with classification-promoted hashing networks.} In addition, we compare with some other hashing methods, especially the networks combined with classification tasks, to illustrate the characteristics of different frameworks. According to the original paper, CFH \cite{ma2020correlation}, sRLH \cite{9638382}, and FISH \cite{chen2022fine} are equipped with ResNet-18, thus we adopt the same backbone. Figure~\ref{fig5} presents the mAP results of different methods. The coarse-grained networks DSH \cite{liu2016deep} and HashNet \cite{cao2017hashnet} show limited discriminative power for describing fine-grained objects, which only obtain 31.87\% and 23.20\% mAP on 12 hash bits. Compared with the networks solely trained by hash learning, i.e., DSaH \cite{jin2020deep}, ExchNet, and SEMICON, our AGMH still exhibits outstanding performance as in Table~\ref{tab1}. 

When it comes to the methods promoted by classification tasks, they can achieve extremely high accuracy on short codes, while the mAP results increase slightly with the extension of hash bits. Different from the hash learning driven by preserving pairwise similarities of hash codes, classification loss directly provides category information during training. Consequently, CFH, sRLH, and FISH can learn refined feature representations for fine-grained objects without hash learning. However, taking the most advanced method FISH as an example, the mAP results drop by 29.44\%, 14.04\%, 11.65\%, and 9.27\% on various hash bits without the assistance of classification losses. It can be inferred that the same situation will occur on CFH and sRLH. In practice, the parameters of sRLH are initialized by minimizing the classification loss on the training set. This paper intends to explore fine-grained image retrieval without category information, thus the classification loss is not added. Although without the improvement of a classification task, AGMH still obtains the best results on 48-bit and 64-bit experiments on CUB200-2011, which exceeds FISH by 2.09\% and 3.59\% respectively.

Furthermore, we simply introduce a classification loss in the learning of AGMH and compare its performance with sRLH in Table~\ref{tab2}. The category information effectively promotes AGMH by 9.55\% and 4.0\% in the 16-bit and 32-bit experiments on CUB200-2011. In particular, AGMH achieves 6.26\%, 6.15\%, 6.01\%, and 5.58\% improvements over sRLH in the experiments of various hash codes on Stanford Dogs, which also illustrate the advantage of our framework. It can be inferred that the category information will further boost AGMH with elaborate optimization, but it is not our focus.

\begin{table}[t]
\caption{The mAP results comparison between AGMH$^{\S}$ and sRLH (based on ResNet-18). $\S$ indicates that AGMH is trained with additional classification loss.}
\label{tab2}
\begin{tabular}{c|c|cccc}
\toprule
\multirow{2}*{Datasets} & \multirow{2}*{Methods} & \multicolumn{4}{c}{\# bits} \\
& & \makebox[0.045\textwidth][c]{16} & \makebox[0.045\textwidth][c]{32} & \makebox[0.045\textwidth][c]{48} & \makebox[0.045\textwidth][c]{64} \\
\hline
\multirow{2}*{CUB200-2011} & sRLH & \textbf{62.68} & 69.37 & 71.27 & 71.60 \\
& AGMH$^{\S}$ & 59.68 & \textbf{76.71} & \textbf{80.73} & \textbf{81.43} \\
\hline
\multirow{2}*{Stanford Dogs} & sRLH & 65.27 & 72.32 & 74.36 & 75.33 \\
& AGMH$^{\S}$ & \textbf{71.53} & \textbf{78.47} & \textbf{80.37} & \textbf{80.91} \\
\bottomrule   
\end{tabular}
\end{table}

\subsection{Ablation Studies}

\begin{table*}[t]
\caption{The mAP results of AGMH with incremental components. “Base” denotes the base hashing network with multiple descriptors to group fine-grained attributes.}
\label{tab3}
\begin{tabular*}{0.95\linewidth}{ccc|cccc|cccc|cccc}
\toprule
\multirow{2}*{Base} & \multirow{2}*{ADL} & \multirow{2}*{SIEA} & \multicolumn{4}{c|}{CUB200-2011} & \multicolumn{4}{c|}{Aircraft} & \multicolumn{4}{c}{Food101} \\
& & & \makebox[0.045\textwidth][c]{12} & \makebox[0.045\textwidth][c]{24} & \makebox[0.045\textwidth][c]{32} & \makebox[0.045\textwidth][c]{48}
& \makebox[0.045\textwidth][c]{12} & \makebox[0.045\textwidth][c]{24} & \makebox[0.045\textwidth][c]{32} & \makebox[0.045\textwidth][c]{48}
& \makebox[0.045\textwidth][c]{12} & \makebox[0.045\textwidth][c]{24} & \makebox[0.045\textwidth][c]{32} & \makebox[0.045\textwidth][c]{48} \\
\hline
\ding{51} & & & 51.59 & 77.00 & 79.96 & 83.06 & 65.97 & 81.22 & 83.26 & 83.10 & 58.81 & 79.87 & 81.13 & 81.27 \\
\ding{51} & \ding{51} & & 51.51 & 76.63 & 80.86 & 83.32 & 66.69 & 82.65 & 83.27 & 83.10 & 57.26 & 80.19 & 80.79 & 80.74 \\
\ding{51} & \ding{51} & \ding{51} & \textbf{56.42} & \textbf{77.44} & \textbf{81.95} & \textbf{83.69} & \textbf{71.64} & \textbf{83.45} & \textbf{83.60} & \textbf{84.91} & \textbf{62.59} & \textbf{80.94} & \textbf{82.31} & \textbf{83.21} \\
\bottomrule   
\end{tabular*}
\end{table*}

\begin{figure}[t]
\centering
\subfigure{\includegraphics[width=1.65in]{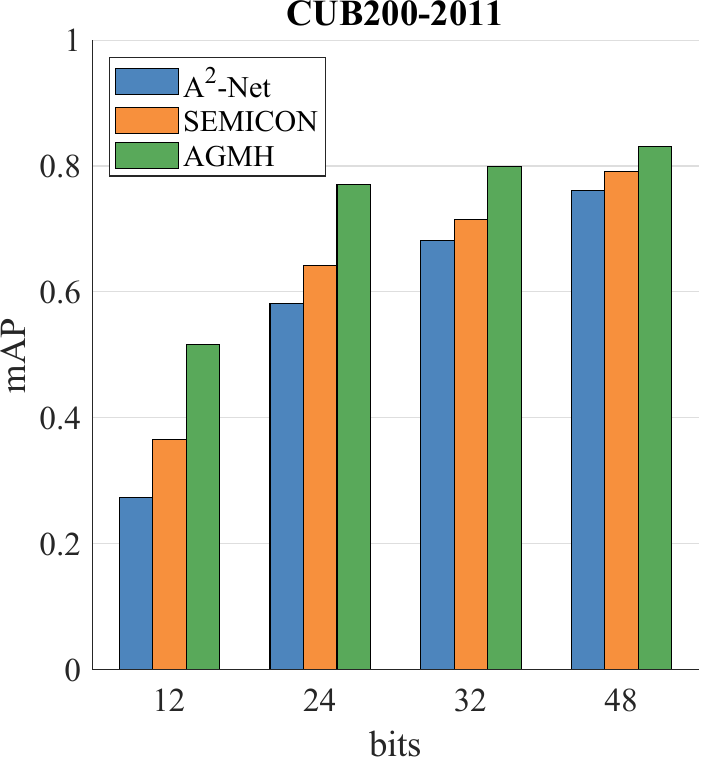}}
\subfigure{\includegraphics[width=1.65in]{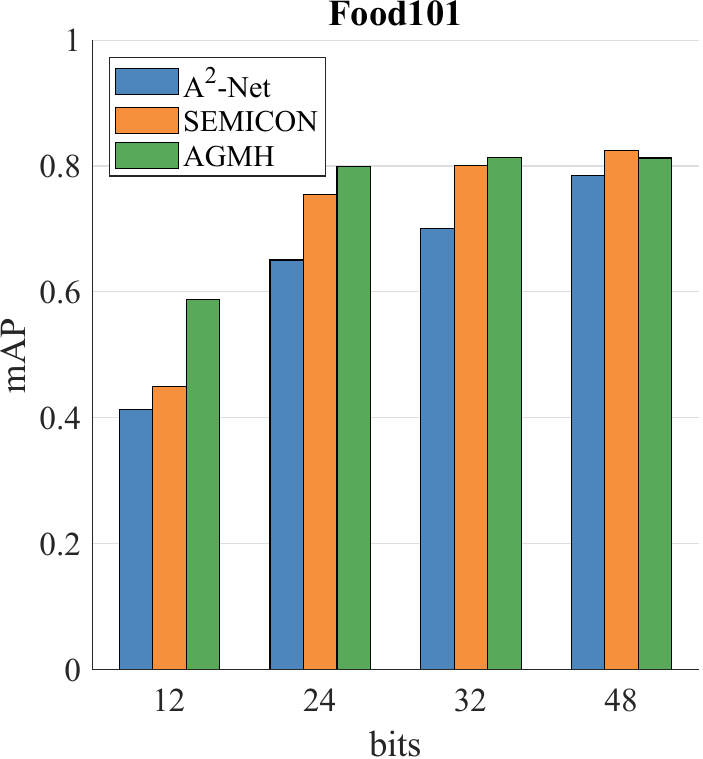}}
\caption{The retrieval accuracy of hash codes generated from the attribute descriptors of AGMH and the attention-guided features of A$^2$-Net and SEMICON.}
\label{fig6}
\end{figure}

\noindent\textbf{Effectiveness of each module.} In this section, we conduct ablation studies on CUB200-2011, Aircraft, and Food101 to verify the effectiveness of each module. The modules are applied incrementally on a base network. As is illustrated in Table~\ref{tab3} and Figure~\ref{fig6}, the base network with attributes grouping method can achieve high mAP results in fine-grained retrieval on various hash bits. Actually, we do not introduce downsampling to the last stage of vanilla backbone, which provides more detailed information for fine-grained attributes and is crucial to strengthen retrieval performance. When we mine attributes without connecting SIEA, the performance of AGMH has no improvement but even declines slightly in some cases, e.g., a 0.37\% drop of 24-bit experiment on CUB200-2011 and a 0.53\% drop of 48-bit experiment on Food101. The introduction of SIEA can mitigate the side effect of attention dispersion loss on hash learning, which further improves retrieval accuracy.

\begin{table}[t]
\caption{Experiments of AGMH with different attention mechanisms. SA: self-attention. EA: external attention. MEA: multi-head external attention. SIEA: stepwise interactive external attention. MEA$^{\dag}$ adopts independent memory for each head while no interaction between tokens.}
\label{tab4}
\begin{tabular}{c|c|ccccc}
\toprule
\makebox[0.09\textwidth][c]{Datasets} & \makebox[0.04\textwidth][c]{\# bits} & SA & EA & MEA & MEA$^{\dag}$ & SIEA \\
\hline  
\multirow{4}*{CUB200-2011} & 12 & 52.62 & 52.03 & 51.92 & 53.37 & \textbf{56.42} \\
 & 24 & 77.44 & 76.33 & 76.56 & 76.39 & \textbf{77.44} \\
 & 32 & 81.55 & 81.01 & 81.23 & 80.67 & \textbf{81.95} \\
 & 48 & 83.41 & 83.31 & 83.17 & 82.87 & \textbf{83.69} \\
\hline
\multirow{4}*{Aircraft} & 12 & 71.01 & 71.15 & 69.46 & 69.06 & \textbf{71.64 }\\
 & 24 & 82.41 & 82.10 & 82.83 & 82.65 & \textbf{83.45} \\
 & 32 & 82.43 & 83.01 & \textbf{83.93} & 83.58 & 83.60 \\
 & 48 & 83.21 & 82.73 & 83.54 & 83.44 & \textbf{84.91} \\
\bottomrule   
\end{tabular}
\end{table}

\noindent\textbf{Compare attribute descriptors with attention-guided features.} To demonstrate the effectiveness of attributes grouping, we compare AGMH with the representative hashing methods A$^2$-Net and SEMICON which are equipped with attention guidance. The hash codes are directly learned from attribute descriptors and attention-guided features, while the additional modules in the frameworks are removed. As can be seen in the bar plots of Figure~\ref{fig6}, the proposed AGMH is superior to other methods on various hash bits and datasets. In particular, AGMH improves the mAP of 12-bit experiments from 36.58\% to 51.59\% on CUB200-2011 and from 44.95\% to 58.81\% on Food101, which benefits from the excellent feature representations of multiple descriptors.

The number of descriptors is set as 6. In our studies, we find that a few descriptors can not represent the diverse fine-grained attributes well, while excessive descriptors lead to information redundancy, which will all degrade model performance.

\noindent\textbf{Comparison with different attntion mechanisms.} Moreover, we employ various attention mechanisms in AGMH to prove the advantages of SIEA in fine-grained attributes mining. For fair comparisons, the number and channel dimension of heads of MEA and MEA$^{\dag}$ are consistent with SIEA. It can be observed in Table~\ref{tab4} that SIEA is more suitable for our framework with the highest mAP results on CUB200-2011 and Aircraft. Compared with external attention, self-attention does not show an obvious advantage as expected. On extremely short hash bits, MEA is slightly inferior to other methods, while its accuracy goes ahead on the long bits, which achieve the highest accuracy of 83.93\% in 32-bit experiment on Aircraft. To MEA$^{\dag}$, its overall performance is worse than MEA, which is in accord with the conclusion of orirginal paper \cite{guo2022beyond}. The attention mechanism does not introduce extra costs in the actual application. As is shown in Table~\ref{tab5}, the retrieval speed will drop to 3ms per image when we input multiple images in a batch.

\begin{table}[t]
\caption{Time consumption of AGMH (based on ResNet-50) with 48-bit hash codes to retrieve a single image from Food101 on one RTX 2080 GPU.}
\label{tab5}
\begin{tabular}{c|cccc}
\toprule
batch size & 1 & 4 & 16 & 64 \\
\hline
speed (ms per image) & 12.68 & 4.35 & 3.33 & 3.03 \\
\bottomrule
\end{tabular}
\end{table}

Overall, the comparisons of retrieval accuracy demonstrate the advance of our proposed framework and its promising practicality in large-scale fine-grained image retrieval.

\section{Conclusion}
In this paper, we propose a novel hashing framework for fine-grained image retrieval, namely Attributes Grouping and Mining Hashing (AGMH). Aiming at the limitations of existing hashing networks with attention guidance, we generate multiple attribute descriptors to enhance the representation of fine-grained objects. An elaborate attention dispersion loss is developed to guide the attention of discrete object attributes. In addition, we design a stepwise interactive external attention to dig into the critical attributes implied in each descriptor and alleviate the disturbance of ADL to hash learning, which does not introduce additional costs in the testing phase. Extensive experiments on five datasets demonstrate that AGMH achieves state-of-the-art performance in fine-grained image retrieval. It is worth noting that AGMH is a simple hash learning framework without additional classification loss, and we believe that it will achieve better results with more supplementary modules and learning tasks.

\begin{acks}
This work is supported by the National Natural Science Foundation of China under grant 62271143. On computing resources, this work is supported by the Big Data Center of Southeast University.
\end{acks}

\newpage

\bibliographystyle{ACM-Reference-Format}
\balance
\bibliography{AGMH}

\end{document}